\documentclass[]{scrartcl}

\usepackage{algorithm}
\usepackage{algorithmic}
\usepackage{amsmath,amssymb}
\usepackage{graphicx}

\title{Methods for Characterizing the Epigenetic Attractors Landscape Associated with Boolean Gene Regulatory Networks}

\author{Jose Davila-Velderrain\textsuperscript{1,2,*}, Luis Juarez-Ramiro\textsuperscript{3}\\Juan C. Martinez-Garcia\textsuperscript{3}, Elena R. Alvarez-Buylla\textsuperscript{1,2,*} \\}

\date{}

\begin{document}

\maketitle

{\raggedright
	
	{\footnotesize	
		\bf{1} Instituto de Ecolog\'ia, Universidad Nacional Aut\'onoma de M\'exico, Cd. Universitaria, M\'exico, D.F. 04510, M\'exico \\
		\bf{2} Centro de Ciencias de la Complejidad (C3), Universidad Nacional Aut\'onoma de M\'exico, Cd. Universitaria, M\'exico, D.F. 04510, M\'exico \\
		\bf{3} Departamento de Control Autom\'atico, Instituto Polit\'ecnico Nacional, A. P. 14-740, 07300 M\'exico, DF, M\'exico
	}
}

\begin{abstract}
Gene regulatory network (GRN) modeling is a well-established theoretical framework for the study of cell-fate specification during developmental processes. Recently, dynamical models of GRNs have been taken as a basis for formalizing the metaphorical model of Waddington’s epigenetic landscape, providing a natural extension for the general protocol of GRN modeling. In this contribution we present in a coherent framework a novel implementation of two previously proposed general frameworks for modeling the {\em Epigenetic Attractors Landscape} associated with boolean GRNs: the {\em inter-attractor} and {\em inter-state} transition approaches. We implement novel algorithms for estimating inter-attractor transition probabilities without necessarily depending on intensive single-event simulations. We analyze the performance and sensibility to parameter choices of the algorithms for estimating inter-attractor transition probabilities using three real GRN models. Additionally, we present a side-by-side analysis of downstream analysis tools such as the attractors' temporal and global ordering in the EAL. Overall, we show how the methods complement each other using a real case study: a cellular-level GRN model for epithelial carcinogenesis. We expect the toolkit and comparative analyses put forward here to be a valuable additional resource for the systems biology community interested in modeling cellular differentiation and reprogramming both in normal and pathological developmental processes.
\end{abstract}

\section*{Introduction}

The postulation of experimentally grounded gene regulatory network (GRN) dynamical models, their qualitative analysis and dynamical characterization in terms of control parameters, and the validation of GRN predictions against experimental observations has become a well-established framework in systems biology -- see, for example: \cite{mendoza1998dynamics, espinosa2004gene, huang2007bifurcation, davila2015descriptive}. There are multiple tools available for the straightforward implementation and analysis of dynamical models of GRNs \cite{Azpeitia2014FlowerDev}. These models are well-suited for the study of cell-fate specification during developmental processes. More recently, dynamical models of GRNs have been taken as a basis for formalizing a century-old developmental metaphor: Waddington's epigenetic landscape \cite{waddington1957strategy, alvarez2008floral, huang2012molecular, Villarreal2012, davila2015reshaping}. The present authors recently introduced the term {\em Epigenetic Attractors Landscape (EAL)} in order to distinguish this modern view of the EL from its metaphorical counterpart (see \cite{davila2015modeling}). Accordingly, here we will refer as EAL to a group of dynamical models grounded in dynamical systems theory and which operationally define an underlying EL associated with GRN dynamics. In this contribution we focus on the EAL associated with the discrete-time boolean description of GRNs grounded on experimental data. \\

Despite growing interest in modeling the EAL, as evidenced by recent model proposals in the study of stem cell differentiation \cite{li2013quantifying} and reprogramming \cite{wang2014epigenetic}, as well as the study of carcinogenesis \cite{wang2014quantitative, zhu2015endogenous} and cancer therapeutics \cite{choi2012attractor, wang2013therapeutic}; unlike the case of GRNs, there are no available tools for the straightforward implementation of EAL models. Furthermore, different EAL models have not been compared directly through side-by-side analysis of the same biological system. This has arguably precluded the wide-spread applicability of EALs. \\

One of the first methodological frameworks proposed to explore the EAL associated with a Boolean GRN was presented by Alvarez-Buylla and collaborators \cite{alvarez2008floral}. Briefly, in its original form this framework rests on three steps: (1) introducing stochasticity into the boolean dynamics by means of the so-called stochasticity in nodes model (SIN), (2) estimating an {\em inter-attractor} transition probability matrix by simulation, and (3) analyzing the temporal evolution of the probability distribution over attractor states (see methods). For the purpose of this contribution, we refer to such framework as the {\em inter-attractor} transition approach (IAT). Recently, a related framework was presented by Zhou and his collaborators \cite{Zhou2014Discrete}. The main differences between this and the former method are: the latter (1) precludes simulation by introducing stochasticity directly into a deterministic transition matrix, and (2) it is based on the estimation of a {\em inter-state} transition probability matrix. We refer to this latter framework as the {\em inter-state} transition approach (IST). Additionally, Zhou and collaborator introduced the idea of a global ordering of attractors in the EAL defined by analyzing the relative stability of attractor states \cite{zhou2014relative}, where stability is quantified in terms of the strength of the attractors  (phenotypes) to endure stochastic disturbances. \\

In this contribution we present in a coherent framework a novel implementation of the two methodologies, as well as associated analysis tools such as the global ordering of the attractors based on relative stabilities, the computation of a quasi-potential landscape based on an stationary probability distribution, and additional tools for downstream analyses and plotting. We use the popular R statistical programming environment (www.R-project.org). For the first framework (IAT), we implement novel algorithms for estimating {\em inter-attractor} transition probabilities without necessarily depending on intensive single-event simulations. For both frameworks (IAT and IST) we exploit the vector-based programming capability of the R language. We analyze the performance and sensibility to parameter choices of the algorithms for estimating {\em inter-attractor} transition probabilities using three GRN models: the Arabidopsis (1) root stem cell niche \cite{azpeitia2010single} and (2) early flower development \cite{davila2015reshaping} GRNs; and (3) a cellular-level GRN model for epithelial carcinogenesis. Additionally, for the latter model we present for the first time, a side-by-side analysis of the two frameworks and show how the methods complement each other. Importantly, we show that the attractor time-ordered transitions obtained by directly estimating an inter-attractor transition matrix are consistent with the global ordering of the attractors obtained by means of their corresponding relative stabilities. All the necessary codes for applying the methods and examples showed herein are made publicly available (see methods below); we expect this toolkit to be a valuable additional resource for the systems biology community.

\section*{Results}

\subsection*{Characterizing the Epigenetic Attractors Landscape}

In this work we organize previously existing, yet dispersed, mathematical analyses into a coherent framework for the characterization of EAL associated with discrete-time boolean description of GRNs grounded on experimental data. Figure 1 schematically represents a general work flow for such characterization. The work flow is supposed to be applicable to an already available and validated experimentally grounded Boolean GRN model (see \cite{Azpeitia2014FlowerDev}). The first necessary step (Fig. 1a) consists of characterizing the state-space associated with the GRN in terms of the attained attractors and their basins, a standard practice in the dynamical analysis of Boolean GRNs (see methods). The second main step consists on estimating either a inter-attractor or inter-state transition probability matrix (or both) (Fig. 1b). The former is the main mathematical structure for the IAT aproach, and the latter for the IST approach (see methods). Downstream analyses of the underlying EAL such as the temporal-order of attractor attainment, the attractor relative stability and global ordering, and the construction of a probabilistic landscape are based on the transition matrices and can be applied afterwards (Fig. 1c).

\subsection*{Inter-attractor Transitions}

A first necessary step in order to explore the EAL associated with a Boolean GRN using the IAT approach is to calculate the probabilities of transition from one attractor to another. In this contribution we present two algorithms for such task (see methods). Algorithm 1 implements what we will refer to as an intuitive mapping-guided random walk in state space. The reasoning is as follows. An initial state is taken at random, which is then mapped to a next state using the stochastic mapping in Equation (3). The basins corresponding to the two states are recorded in order. Subsequently, another state is picked at random from the latter basin, and the mapping procedure is repeated. The procedure is repeated $Nsteps$ number of times, each time taking at random a state from the present basin, and the goal is to record a stochastic realization of the transitions from one basin to another. Algorithm 2, on the other hand, considers all the possible states, repeats them $Nreps$ number of times in a single data structure, and maps them using Equation (3) as well (for details, see methods). An important technical issue is then how to select the parameters $Nsteps$ and $Nreps$, respectively. Specially because this type of simulation approaches have been qualified as requiring large number of time-consuming sampling \cite{Zhou2014Discrete}. \\

For each algorithm we tested how the estimate of the inter-attractor transition matrix changes as the parameter value increases. We used three real GRN models for testing: {\em Arabidopsis} single-cell root stem cell niche GRN (root-GRN) \cite{azpeitia2010single}, {\em Arabidopsis} floral organ determination GRN (flower-GRN) \cite{Azpeitia2014FlowerDev}, and a cellular-level GRN model for epithelial carcinogenesis (cancer-GRN). We found that for models of size common to GRN developmental modules (i.e., $8-15$ genes) the estimation obtained with small values of the parameter rapidly converges to that obtained by using large values (e.g., $\approx 10^6$). Figure 2 shows how the distance between the estimate obtained using a value $Nsteps(Nreps) = i$ and that obtained using  $Nsteps=10^6 $ and $Nreps=10^3$ for Algorithms 1 and 2, respectively. These results correspond to the three GRN models: root (Fig. 2a-b), cancer (Fig. 2c-d), and flower (Fig. 2e-f). Additionally, we show that the estimate obtained with one of the algorithms also rapidly converges to that obtained with the other algorithm. Figure 3 shows how the distance between the estimate obtained using one algorithm with a parameter value $i$ and that obtained using the other algorithm with a large parameter value decreases as $i$ increases. Based on this latter analysis we conclude that, for GRNs of sizes $8-15$ genes, using a value of the order of $Nsteps = 10^4$ for algorithm 1 and $Nreps=10^2$ would be sufficient to achieve an accuracy similar to that achieved using large values (i.e, $10^6$ and $10^3$, respectively), decreasing then the involved computational cost.

\subsection*{Characterizing the EAL}

In this section we provide as an example the analysis of the EAL underlying a cellular-level GRN model for epithelial carcinogenesis. The details of the construction and validation of such network model are being published by the authors elsewhere. The GRN comprises 9 main regulators of epithelial carcinogenesis (Fig. 4), and its dynamical characterization uncovers 3 fixed-point attractor corresponding to the epithelial, senescent, and mesenchymal stem-like cellular phenotypes. We applied the two approaches (IAT and IST) to the cancer-GRN, and for the IAT approach we applied the two algorithms proposed herein. Accordingly, we estimated two inter-attractor transition matrices and one inter-state transition matrix. For simplicity in all cases we kept fixed a single value for the error parameter $\xi = 0.05$. Using the estimated matrices, we applied the downstream analyses depicted in Figure 1c. Figure 5 shows two graphs plotting the temporal evolution of the occupation probability distribution over attractor states epithelial (black), senescent (red) and mesenchymal (green) -- conditioned on an initial distribution where all the cellular population is in the epithelial attractor state. The uncovered attractor time-order is indicated by sequential vertical lines: the order is epithelial $\rightarrow$ senescent $\rightarrow$ mesenchymal. Importantly, the two algorithms give the same qualitative result. \\

Subsequently, we uncovered the global ordering of attractors by calculating the relative stabilities and net transition rates between pairs of attractors using the two inter-attractor transitions estimated with the two algorithms (for details, see methods). Figure 6 shows the plot of two graphs where an arrow appears in color red if the calculated transition rate between the attractor is positive in the indicated direction. The global ordering corresponds to the path comprised by directed arrows passing by the three attractors, here: epithelial $\rightarrow$ senescent $\rightarrow$ mesenchymal. Thus, the global ordering is consistent with the attractor time-order, as long as the latter is conditioned on having the total probability mass in the epithelial attractor as initial state. Again, the two algorithms  produce the same qualitative result. \\

Finally, we used the estimated inter-state transition matrix obtained with the IST approach to derive a graphical probabilistic landscape (see methods). The landscape is based on the stationary probability distribution $\mathbf{u}_{ss}$ obtained by numerical simulation (see methods). Figure 7 and 8 show a 3D-surface and a contour plot respectively. The graphical landscape was derived by first mapping all the state vectors in the sate-space into a low dimensional space by the dimensionality reduction technique principal component analysis. The first two component are taken as the coordinates in the 3D plot, where the z-coordinate corresponds to the values $-log(\mathbf{u}_{ss})$. The surface is inferred by interpolating the spaced data points using the technique of thin plate spline regression \cite{furrer2009fields}. The 3D-surface plot nicely shows the relative stability of the states by means of their probability, the lower states display a higher level of relative stability than the states initially located at higher places. The route from the attractors of less stability to that with the highest consists with the global ordering uncovered above. However, in the case of the IST transition and the probabilistic landscape we have additional information concerning the relative stability of all the transitory states in state space.

\section*{Discussion}

Boolean GRN models are well-established tools for the mechanistic study of the establishment of cellular phenotypes during developmental dynamics. Their simplicity and deterministic nature are well-suited for answering questions regarding the sufficiency of molecular players and interactions necessary to explain observed cellular phenotypes. In the present contribution we present methods to study an extended Boolean GRN model which takes stochasticity into consideration, necessary for studying cell-state transition events.\\

In the case of the stochastic Boolean GRNs, the model of interest involves random samples with a non-trivial dependence structure. In such cases, efficient simulation algorithms are needed in order to explore and characterize the underlying structure and to understand the behavioral (dynamical) consequences of the constrains imposed by such structure. Accordingly, we propose two algorithms of general applicability, and show how these can be used to estimate transitions probabilities in an efficient way from moderate size GRNs similar to those proposed as developmental modules driving developmental processes. Although we show that the two algorithms generate consistent estimates, one or the order may be preferred depending on the GRN in question, as well as the computational resources at hand. Algorithm 1 is likely to be preferred in the case of larger GRNs, as it is not constrained by the size of the GRN per se, but the number of steps chosen in the simulation. On the other hand, given the declarative representation used in Algorithm 2, its performance is constrained by the memory available. Algorithm 2, however, may be preferred for fast estimates in small to moderate size GRNs ($< 15$ genes). Importantly, although we tested the performance of the algorithms in terms of the number of steps chosen for the simulations, the results should not be generalized without caution given that we only used three real GRNs, and the results may vary either for larger GRNs or state spaces with more complex structures. \\

For illustrative purposes we applied all the methods and downstream analyses presented herein to a specific GRN: a cellular-level GRN model for the description of the phenotypic transitions involved in epithelial carcinogenesis. We show that for this case, the uncovered temporal-order of attractor attainment is consistent with the global ordering based on the exploration of the dynamics of the relative stability of the uncovered attractors, both calculated from a inter-attractor transition probability matrix. The result of the former is conditioned on the initial occupation probability taken. An interesting open problem would be to generalize this relationship using GRNs with divers structures, for example to ask if the global ordering of attractors is robust enough as to drive most initial distributions into a consistent temporal ordering. An additional interesting questions would be, what does this relationship tells us about the structural constraints imposed by the GRN. The tools and implementation presented here may prove useful for such theoretical studies.       \\

Finally, we present tools for deriving a probabilistic landscape from an estimated inter-state transition matrix in terms of the stationary probability distribution over state space. This latter analysis and the associated graphical tools can be applied to systematically study how the system responds to perturbations resulting in a reshaped EAL. Structural alterations of the EAL may predict the induction of preferential cell-state transitions such as the case of reprogramming strategies \cite{zhou2011understanding} or therapeutic interventions against the stabilization of a cancer attractor \cite{huang2013escape, wang2013therapeutic}.      \\

Overall, in this contribution we present in a coherent framework a novel implementation of general frameworks for modeling the {\em Epigenetic Attractors Landscape} associated with boolean GRNs. We provide analysis of the method performance and show how they can be applied to real case GRNs. We expect the toolkit and comparative analyses put forward here to be a valuable additional resource for the systems biology community interested in modeling cellular differentiation and reprogramming both in normal and pathological developmental processes.

\section*{Materials and Methods}

\subsection*{Boolean Gene Regulatory Networks}

A Boolean network models a dynamical system assuming both discrete time and discrete state variables. This is expressed formally with the mapping:

\begin{equation}
x_i(t+1) = F_i(x_1(t),x_2(t),...,x_k(t)),
\end{equation}

where the set of functions $F_i$ are logical propositions (or truth tables) expressing the relationship between the genes that share regulatory interactions with the gene $i$, and where the state variables $x_i(t)$ can take the discrete values $1$ or $0$ indicating whether the gene $i$ is expressed or not at a certain time $t$, respectively. \\

A completely specified Boolean GRN model is analyzed by either of two methods: (1) by exhaustive computational characterization of the state space in terms of attained attractors and their basins of attractions (used in IAT), or (2) by defining a matrix explicitly encoding the mapping in Equation (1) (used in IST). Specifically, for the latter method, following \cite{zhou2014relative} the mapping in Equation (1) is used to define a single-step $2^n \times 2^n$ transition matrix $\mathbf{T}$ with elements $t_{i,j}$, where:

\begin{equation}
t_{i,j} = 
\left\{
\begin{aligned}
& 1,     && \mathbf{x}_j = \mathbf{F}(\mathbf{x}_i) \\
& 0,      && Otherwise.
\end{aligned}
\right.
\end{equation}

Here $\mathbf{x}_i$ is the network state $i$ from the state-space of size $2^n$ corresponding to a network of $n$ genes, and $\mathbf{F}$ represents the vector of $n$ functions represented element-wise in Equation (1). Given the deterministic character of the mapping in Equation (1), the matrix $\mathbf{T}$ is sparse, each row $i$ having only one element where $t_{i,j}=1$. The matrix $\mathbf{T}$ constitutes a declarative representation which includes the complete information of the mapping in Equation (1): the matrix $\mathbf{T}$ assign to each of the states $\mathbf{x}_k$, where $k \in \{1,...,2^n\}$, its corresponding state in time $t+1$.

\hfil


\subsection*{Inter-Attractor Transition Approach}

\subsubsection*{Including Stochasticity}

Following \cite{alvarez2008floral, Azpeitia2014FlowerDev, davila2015modeling}, a Boolean GRN is extended into a discrete stochastic model by means of  the so--called stochasticity in nodes (SIN) model. In this model, a constant probability of error $\xi$ is introduced for the deterministic Boolean functions as follows:

\begin{equation}
\begin{aligned}
&  P_{x_i(t+1)}[F_i(\mathbf{x}_{reg_i}(t))] = 1- \xi, \\
& P_{x_i(t+1)}[1 - F_i(\mathbf{x}_{reg_i}(t))] = \xi.
\end{aligned}
\end{equation}

It is assumed that the probability that the value of the random variable $x_i(t+1)$ (a gene) is determined or not by its associated logical function $F_i(\mathbf{x}_{reg_i}(t))$ is $1- \xi$ or $\xi$, respectively. The probability $\xi$ is a scalar constant parameter acting independently per gene. The vector $\mathbf{x}_{reg_i}$ represents the regulators of gene $i$.

\subsubsection*{Inter-Attractor Transition Probability Estimation}

An attractor transition probability matrix $\Pi$ with components: 
\begin{equation}
\pi_{ij} = P(A_{t+1}=j|A_t=i),
\end{equation}
representing the probability that an attractor $j$ is reached from an attractor $i$ is estimated by either of two simulation-based algorithms proposed herein (see results).

\begin{algorithm}
	\caption{Simulate {\em inter-attractor} stochastic realization}
	\begin{algorithmic} 
		\STATE {\bf Initiate} storage[{\em Nsteps}]
		\STATE from $state \, space=\{1,...,2^n\}$ pick randomly initial state $\mathbf{x}_i$
		\STATE storage[1] $\leftarrow$ basin $k$ $\leftarrow$  map $\leftarrow$  $\mathbf{x}_i$
		
		\FOR{({\em stepN} in 2 to {\em Nsteps})} 
		\STATE state $\mathbf{x}_j$ $\leftarrow$   stochastic mapping Eq(2) $\leftarrow$ state $\mathbf{x}_i$
		\STATE storage[stepN] $\leftarrow$ basin $k$ $\leftarrow$  map $\leftarrow$  $\mathbf{x}_j$
		\STATE from $sub \, space = \{basin \, k\}$  pick randomly state $\mathbf{x}_i$
		\ENDFOR
		\STATE {\bf return} storage
	\end{algorithmic}
\end{algorithm}

\begin{algorithm}
	\caption{Implicit bit-flip simulation}
	\begin{algorithmic} 
		\STATE {\bf Initiate} storage $j \times j$ matrix $\Pi$, $j \in \{1, ..., n_{attractors} \}$
		\STATE Generate $state \, space=\{\mathbf{x}_1,...,\mathbf{x}_{2^n}\}$
		\STATE Generate set $\mathbf{X_{t+1}} = \mathbf{F}(state \, space)$
		\STATE $\mathbf{X_{t+1}^{pert}}$ $\leftarrow$  repeat $\mathbf{X_{t+1}}$ element-wise $Nsteps$ times  
		\STATE Generate perturbation indicator vector $\mathbf{piv}$:
		\STATE $\, \, \,$ $\mathbf{piv}$ $\leftarrow$ simulate $Nsteps \times n \times 2^n$ observations from $Bin(n=1,\xi)$
		
		\FOR{$\mathbf{piv}[i]=1$} 
		\STATE Apply error in $\mathbf{X_{t+1}^{pert}}[i]$ , $i \in \{1,...,Nsteps \times n \times 2^n \}$ 
		\ENDFOR
		
		\STATE $\mathbf{X^{pert}} \leftarrow$ split $\mathbf{X_{t+1}^{pert}}$ in $n$-size state vectors $\mathbf{x}_k, k \in \{1,...,Nsteps \, \times \, 2^n\}$
		
		\FOR {each $\mathbf{x}_i$ in $state \, space$} 
		\STATE basin $j$ $\leftarrow$  map $\mathbf{x}_i \,$ 
		\ENDFOR
		
		\FOR {each $\mathbf{x}_k$ in $\mathbf{X^{pert}}$} 
		\STATE basin $j$ $\leftarrow$  map $\mathbf{x}_k$ 
		\ENDFOR
		
		\STATE update $\pi_{j,j}$
		\STATE {\bf return} storage matrix
	\end{algorithmic}
\end{algorithm}

In Algorithm 2, $Bin(n=1,\xi)$ refers to a binomial distribution given by $Bin(k|n,\xi) = \binom {n} {k} \ \xi^k(1-\xi)^{n-k}$. In the special case used here (with $n=1$) the distribution corresponds to a Bernoulli distribution. Thus, what we call {\em perturbation indicator vector} effectively simulates tossing a biased coin $Nsteps \, \times \, n \, \times \, 2^n$ times. Each outcome $x=1$ indicates the position where an error in the mapping has occurred, according to Equation (3).    \\

The elements $\pi_{ij}$ of the matrix $\Pi$ are obtained as maximum likelihood estimates based on the empirical transition probability resulting from the simulations from either algorithm 1 or 2. 

\subsection*{Inter-State Transition Probability Approach}

\subsubsection*{Including Stochasticity}

For the IST approach, following \cite{Zhou2014Discrete, zhou2014relative}, stochasticity is introduced in a declaractive manner (i.e., by means of a single structure representation) using a binomial distribution. Specifically, the effect of noise on each possible single-state transition is represented by introducing a noise matrix $\mathbf{N}$ with elements

\begin{equation}
N_{i,j} = 
\left\{
\begin{aligned}
& \binom {n} {d_{ij}} \ \xi^{d_{ij}}(1-\xi)^{n-d_{ij}},     && i \neq j \\
& 0,      && i = j
\end{aligned}
\right.
\end{equation}

where $d_{ij}$ is the Hamming distance between the states $i$ and $j$ (i.e., $d_{ij} = \lVert \mathbf{x}_i - \mathbf{x}_j \rVert_H$ ). This representation formalizes an intuitive notion: the effect of noise on the system is more (less) likely to produce a state less (more) similar to the initial state.\\

\subsubsection*{Inter-State Transition Probability Estimation}

A single object including both stochastic perturbations and deterministic mapping is obtained by adding the noise matrix $\mathbf{N}$ and the deterministic single-step transition matrix $\mathbf{T}$ (see Equation 2) as follows    

\begin{equation}
\mathbf{\Pi} = (1-\xi)^n \mathbf{T} + \mathbf{N}
\end{equation}

After normalizing a transition probability matrix $\Pi$ is obtained with components 

\begin{equation}
\pi_{ij} = P(\mathbf{x}_{t+1}=j|\mathbf{x}_t=i).
\end{equation}

The components $\pi_{ij}$ represent the probability that a state $j$ is reached from a state $i$, where $i,j \in \{1, ... 2^n \}$.

\subsection*{Temporal Evolution of States/Attractors Probability}

In both approaches (IAT and IST) a sequence of random variables $\{C_t : t \in \mathbb{N}\}$ is considered a Markov chain (MC). In IAT (IST) $C_T$ takes as values the different attractors (states), the elements $\pi_{i,j}$ representing inter-attactor(states) transition probabilities, and the matrix $\Pi$ the (one-step) transition probability matrix. As the probabilities do not depend on time, the MC is homogeneous. \\

The occupation probability distribution $P(C_t = j)$ -- i.e., the probability that the chain is in state (attractor or state) $j$ at a given time $t$ -- is denoted by the row vector $\mathbf{u}(t)$. The probabilities temporally evolve according to the dynamic equation

\begin{equation}
\mathbf{u}(t+1) = \mathbf{u}(t) \mathbf{\Pi}.
\end{equation}

Taking $\mathbf{u}(0)$ as the initial distribution of the MC, the equation reads $ \mathbf{u}(1) = \mathbf{u}(0) \mathbf{\Pi}.$ By linking the occupation probabilities iteratively we get $ \mathbf{u}(t) = \mathbf{u}(0) \mathbf{\Pi}^t$: the occupation probability distribution at time $t$ can be obtained directly by matrix exponentiation.

\subsection*{EAL Analyses}

\subsubsection*{Temporal-order of Attractor Attainment}

Having obtained the temporal evolution of the occupation probability distribution $\mathbf{u}(t)$ given an initial distribution $\mathbf{u}(0)$ by numerically solving Equation (8); following \cite{alvarez2008floral}, it is assumed that the most likely time for an attractor to be reached is when the probability of reaching that particular attractor is maximal. Therefore, the temporal sequence in which attractors are attained is obtained by determining the sequence in which their maximum probabilities are reached using $\mathbf{u}(t)$.  

\subsubsection*{Probabilistic Landscape}

A stationary probability distribution of a MC is a distribution $\mathbf{u}_{ss}$ which satisfies the steady state equation $\mathbf{u}_{ss} = \mathbf{u}_{ss} \mathbf{\Pi}$. The stationary probability distribution, if exists, is calculated either by solving the equation $\mathbf{u}_{ss}(\mathbf{I}-\mathbf{\Pi})=0$, where $\mathbf{I}$ is the $n \, \times \, n$ identity matrix \cite{wilkinson2011stochastic}; or by numerically solving Equation (8), as $\mathbf{u}_{ss}$ corresponds to the {\em long-run distribution} of the MC: $\mathbf{u}_{ss} = \lim_{t \to \infty} \mathbf{u}(t)$ \cite{bolstad2011understanding}. A probabilistic landscape $U$ -- also called a quasi-potential -- can be obtaining by mapping the distribution $\mathbf{u}_{ss}$ using $-ln(\mathbf{u}_{ss})$. Such landscape reflects the probability of states and it provides a global characterization and a stability measure of the GRN system \cite{wang2015landscape}.

\subsubsection*{Attractor Relative Stability and Global Ordering Analyses}

A relative stability matrix $\mathbf{M}$ is calculated which reflects the transition barrier between any two states based on the mean first passage time (MFPT). The transition barrier in the EAL epitomizes the ease for transitioning from one attractor to another. The ease of transitions, in turn, offers a notion of relative stability. Zhou and collaborators recently proposed that a GRN has a consistent global ordering of all of the attractors which can be uncovered by considering their relative stabilities \cite{Zhou2014Discrete, zhou2014relative}. A net transition rate between attractor $i$ and $j$ is defined in terms of the MFPT as follows:

\begin{equation}
d_{i,j} = \frac{1}{MFPT_{i,j}} - \frac{1}{MFPT_{j,i}}
\end{equation}

The consistent global ordering of the attractors is defined based on the formula proposed in \cite{zhou2014relative}. Briefly, the consistent global ordering of the attractors is given by the attractor permutation in which all transitory net transition rates from an initial attractor to a final attractor are positive. The MFPTs are calculated either by implementing the matrix-based algorithm proposed in \cite{sheskin1995computing} or by means of numerical simulation.

\subsection*{Implementation}

All the methods presented here were implemented using the {\em R} statistical programming environment (www.R-project.org). The code relies on the following packages: {\em BoolNet}, for the dynamical analysis of Boolean networks \cite{mussel2010boolnet}; {\em expm}, for matrix computations \cite{goulet2013expm}; {\em igraph}, for network (graph) analyses \cite{csardi2006igraph}; {\em markovchain} for MC analysis and  inference; and {\em fields}, for surface plotting \cite{furrer2009fields}. The code, including tutorial and examples, is publicly available at {\em https://github.com/JoseDDesoj/Epigenetic-Attractors-Landscape-R}.

\section*{Figure legends}

\textbf{Fig 1. Schematic representations of the general work flow for characterizing the EAL.}
a) The starting point is the dynamical characterization of an experimentally grounded GRN Boolean model in terms of attained attractors and corresponding basins. b) Depending on the downstream analyses of interest, one may proceed by calculating an inter-attractor (IAT) or inter-state (IST) transition matrix, or both. c) Using the calculated transition matrix as input, downstream analysis tools can be applied: the attractor time- and global order from the IAT matrix, and the probabilistic landscape from the IST matrix.    \\

\textbf{Fig. 2 Distance between estimates as a function of parameters $Nsteps$ and $Nreps$. }
The plots show the euclidean distance (y axis) between the estimated transition probabilities using a value $i$ of $Nsteps$ for Algorithm 1 and of $Nreps$ for Algorithm 2 (x axis) and the corresponding estimates using a value of $Nsteps=10^6 $ and $Nreps=10^3$. Plots show calculations for the three GRNs used: root (Fig. 2a-b), cancer (Fig. 2c-d), and flower (Fig. 2e-f).\\

\textbf{Fig. 3 Distance between estimates obtained with Algorithms 1 and 2.}
The plots a, c, and e show the euclidean distance between the estimated transition probabilities obtained using Algorithm 1 with a value $i$ of $Nsteps$ (x axis) and the corresponding estimates obtained using Algorithm 2 with a value of $Nreps=10^3$. Plots b, d, and f show the euclidean distance between the estimated transition probabilities obtained using Algorithm 2 with a value $i$ of $Nreps$ (x axis) and the corresponding estimates obtained using Algorithm 1 with a value of $Nsteps=10^6$. Vertical, dotted lines indicate a tentative minimal value for the corresponding parameter ($Nsteps$ or $Nreps$) able to provide estimates comparable with those obtained using large values. Plots show calculations for the three GRNs used: root (Fig. 2a-b), cancer (Fig. 2c-d), and flower (Fig. 2e-f).  \\

\textbf{Fig 4. Gene regulatory network for epithelial carcinogenesis. }
Nodes represent genes, and arrows represent experimentally characterized interactions. The nature of the interaction (activation or inhibition) is not specified, given that this information is implicit in the logical rules specifying the boolean dynamical model.\\

\textbf{Fig 5. Temporal sequence of cell--fate attainment pattern under the stochastic Boolean GRN model during epithelial carcinogenesis.}
The plots show the maximum probability $P$ of attaining each attractor, as a function of time (in iteration steps). Vertical lines mark the time when maximal probability of each attractor occurs. The most probable sequence of cell attainment is: epithelial(E) $\rightarrow$ senescent(S) $\rightarrow$ mesenchymal. Both algorithms uncover the same time-order pattern.  \\

\textbf{Fig 6. Graph-based representation of attractors transitions. }
Attractor transitions having a positive net transition rate are connected by arrows, which indicate the directionality of the transitions. The global ordering corresponds to the path comprised by directed arrows passing by the three attractors, here: epithelial $\rightarrow$ senescent $\rightarrow$ mesenchymal, resulting in a global probability flow across the EAL.\\

\textbf{Fig. 7 3D-surface (a) and a contour plot (b) representation of the probabilistic landscape.}
The landscape is based on the stationary probability distribution $\mathbf{u}_{ss}$ and was derived by mapping the sate-space into a low dimensional space using principal component analysis. The first two component are taken as x-y coordinates with the corresponding $-log(\mathbf{u}_{ss})$ values as the z-coordinate. The surface is inferred by interpolation.


\bibliographystyle{unsrt}

\bibliography{sample}

\begin{thebibliography}{10}

\bibitem{mendoza1998dynamics}
Luis Mendoza and Elena~R Alvarez-Buylla.
\newblock Dynamics of the genetic regulatory network for arabidopsis thaliana
  flower morphogenesis.
\newblock {\em Journal of theoretical biology}, 193(2):307--319, 1998.

\bibitem{espinosa2004gene}
Carlos Espinosa-Soto, Pablo Padilla-Longoria, and Elena~R Alvarez-Buylla.
\newblock A gene regulatory network model for cell-fate determination during
  arabidopsis thaliana flower development that is robust and recovers
  experimental gene expression profiles.
\newblock {\em The Plant Cell Online}, 16(11):2923--2939, 2004.

\bibitem{huang2007bifurcation}
Sui Huang, Yan-Ping Guo, Gillian May, and Tariq Enver.
\newblock Bifurcation dynamics in lineage-commitment in bipotent progenitor
  cells.
\newblock {\em Developmental biology}, 305(2):695--713, 2007.

\bibitem{davila2015descriptive}
J~Davila-Velderrain, JC~Martinez-Garcia, and ER~Alvarez-Buylla.
\newblock Descriptive vs. mechanistic network models in plant development in
  the post-genomic era.
\newblock {\em Plant Functional Genomics: Methods and Protocols}, pages
  455--479, 2015.

\bibitem{Azpeitia2014FlowerDev}
Eugenio Azpeitia, Jose Davila-Velderrain, Carlos Villarreal, and Elena~R
  Alvarez-Buylla.
\newblock Gene regulatory network models for floral organ determination.
\newblock In {\em Flower Development: Methods and Protocols}. Springer, 2014.

\bibitem{waddington1957strategy}
Conrad~Hal Waddington.
\newblock {\em The strategy of genes}.
\newblock London: George Allen \& Unwin, Ltd., 1957.

\bibitem{alvarez2008floral}
Elena~R Alvarez-Buylla, {\'A}lvaro Chaos, Maximino Aldana, Mariana
  Ben{\'\i}tez, Yuriria Cortes-Poza, Carlos Espinosa-Soto, Diego~A
  Hartas{\'a}nchez, R~Beau Lotto, David Malkin, Gerardo J~Escalera Santos,
  et~al.
\newblock Floral morphogenesis: stochastic explorations of a gene network
  epigenetic landscape.
\newblock {\em Plos one}, 3(11):e3626, 2008.

\bibitem{huang2012molecular}
Sui Huang.
\newblock The molecular and mathematical basis of waddington's epigenetic
  landscape: A framework for post-darwinian biology?
\newblock {\em Bioessays}, 34(2):149--157, 2012.

\bibitem{Villarreal2012}
Carlos Villarreal, Pablo Padilla-Longoria, and Elena~R Alvarez-Buylla.
\newblock {General theory of genotype to phenotype mapping: derivation of
  epigenetic landscapes from N-node complex gene regulatory networks.}
\newblock {\em Physical review letters}, 109(11):118102, September 2012.

\bibitem{davila2015reshaping}
Jose Davila-Velderrain, Carlos Villarreal, and Elena~R Alvarez-Buylla.
\newblock Reshaping the epigenetic landscape during early flower development:
  induction of attractor transitions by relative differences in gene decay
  rates.
\newblock {\em BMC systems biology}, 9(1):20, 2015.

\bibitem{davila2015modeling}
Jose Davila-Velderrain, Juan~C Martinez-Garcia, and Elena~R Alvarez-Buylla.
\newblock Modeling the epigenetic attractors landscape: toward a post-genomic
  mechanistic understanding of development.
\newblock {\em Frontiers in genetics}, 6, 2015.

\bibitem{li2013quantifying}
Chunhe Li and Jin Wang.
\newblock Quantifying cell fate decisions for differentiation and reprogramming
  of a human stem cell network: landscape and biological paths.
\newblock {\em PLoS computational biology}, 9(8):e1003165, 2013.

\bibitem{wang2014epigenetic}
Ping Wang, Chaoming Song, Hang Zhang, Zhanghan Wu, Xiao-Jun Tian, and Jianhua
  Xing.
\newblock Epigenetic state network approach for describing cell phenotypic
  transitions.
\newblock {\em Interface Focus}, 4(3):20130068, 2014.

\bibitem{wang2014quantitative}
Gaowei Wang, Xiaomei Zhu, Jianren Gu, and Ping Ao.
\newblock Quantitative implementation of the endogenous molecular--cellular
  network hypothesis in hepatocellular carcinoma.
\newblock {\em Interface focus}, 4(3):20130064, 2014.

\bibitem{zhu2015endogenous}
Xiaomei Zhu, Ruoshi Yuan, Leroy Hood, and Ping Ao.
\newblock Endogenous molecular-cellular hierarchical modeling of prostate
  carcinogenesis uncovers robust structure.
\newblock {\em Progress in biophysics and molecular biology}, 2015.

\bibitem{choi2012attractor}
Minsoo Choi, Jue Shi, Sung~Hoon Jung, Xi~Chen, and Kwang-Hyun Cho.
\newblock Attractor landscape analysis reveals feedback loops in the p53
  network that control the cellular response to dna damage.
\newblock {\em Science signaling}, 5(251):ra83--ra83, 2012.

\bibitem{wang2013therapeutic}
Wei Wang.
\newblock Therapeutic hints from analyzing the attractor landscape of the p53
  regulatory circuit.
\newblock {\em Science signaling}, 6(261):pe5--pe5, 2013.

\bibitem{Zhou2014Discrete}
Joseph~Xu Zhou, Xiaojie Qiu, Aymeric~Fouquier d'Herouel, and Sui Huang.
\newblock Discrete gene network models for understanding multicellularity and
  cell reprogramming: From network structure to attractor landscapes landscape.
\newblock {\em In: Computational Systems Biology. Second Edition. Elsevier},
  pages 241--276, 2014.

\bibitem{zhou2014relative}
Joseph~Xu Zhou, Areejit Samal, Aymeric~Fouquier d'H{\`e}rou{\"e}l, Nathan~D
  Price, and Sui Huang.
\newblock Relative stability of network states in boolean network models of
  gene regulation in development.
\newblock {\em arXiv preprint arXiv:1407.6117}, 2014.

\bibitem{azpeitia2010single}
Eugenio Azpeitia, Mariana Ben{\'\i}tez, Iliusi Vega, Carlos Villarreal, and
  Elena~R Alvarez-Buylla.
\newblock Single-cell and coupled grn models of cell patterning in the
  arabidopsis thaliana root stem cell niche.
\newblock {\em BMC systems biology}, 4(1):134, 2010.

\bibitem{furrer2009fields}
Reinhard Furrer, Douglas Nychka, and Stephen Sain.
\newblock fields: Tools for spatial data.
\newblock {\em R package version}, 6(11), 2009.

\bibitem{zhou2011understanding}
Joseph~X Zhou and Sui Huang.
\newblock Understanding gene circuits at cell-fate branch points for rational
  cell reprogramming.
\newblock {\em Trends in Genetics}, 27(2):55--62, 2011.

\bibitem{huang2013escape}
Sui Huang and Stuart Kauffman.
\newblock How to escape the cancer attractor: rationale and limitations of
  multi-target drugs.
\newblock In {\em Seminars in cancer biology}, volume~23, pages 270--278.
  Elsevier, 2013.

\bibitem{wilkinson2011stochastic}
Darren~J Wilkinson.
\newblock {\em Stochastic modelling for systems biology}.
\newblock CRC press, 2011.

\bibitem{bolstad2011understanding}
William~M Bolstad.
\newblock {\em Understanding computational Bayesian statistics}, volume 644.
\newblock John Wiley \& Sons, 2011.

\bibitem{wang2015landscape}
Jin Wang.
\newblock Landscape and flux theory of non-equilibrium dynamical systems with
  application to biology.
\newblock {\em Advances in Physics}, 64(1):1--137, 2015.

\bibitem{sheskin1995computing}
Theodore~J Sheskin.
\newblock Computing mean first passage times for a markov chain.
\newblock {\em International Journal of Mathematical Education in Science and
  Technology}, 26(5):729--735, 1995.

\bibitem{mussel2010boolnet}
Christoph M{\"u}ssel, Martin Hopfensitz, and Hans~A Kestler.
\newblock Boolnet—an r package for generation, reconstruction and analysis of
  boolean networks.
\newblock {\em Bioinformatics}, 26(10):1378--1380, 2010.

\bibitem{goulet2013expm}
Vincent Goulet, Christophe Dutang, Martin Maechler, David Firth, Marina
  Shapira, and Michael Stadelmann.
\newblock expm: Matrix exponential.
\newblock {\em R package version 0.99-0}, 2013.

\bibitem{csardi2006igraph}
Gabor Csardi and Tamas Nepusz.
\newblock The igraph software package for complex network research.
\newblock {\em InterJournal, Complex Systems}, 1695(5):1--9, 2006.

\end{thebibliography}


\begin{figure}[p]
	\centering
	\includegraphics[width=150mm]{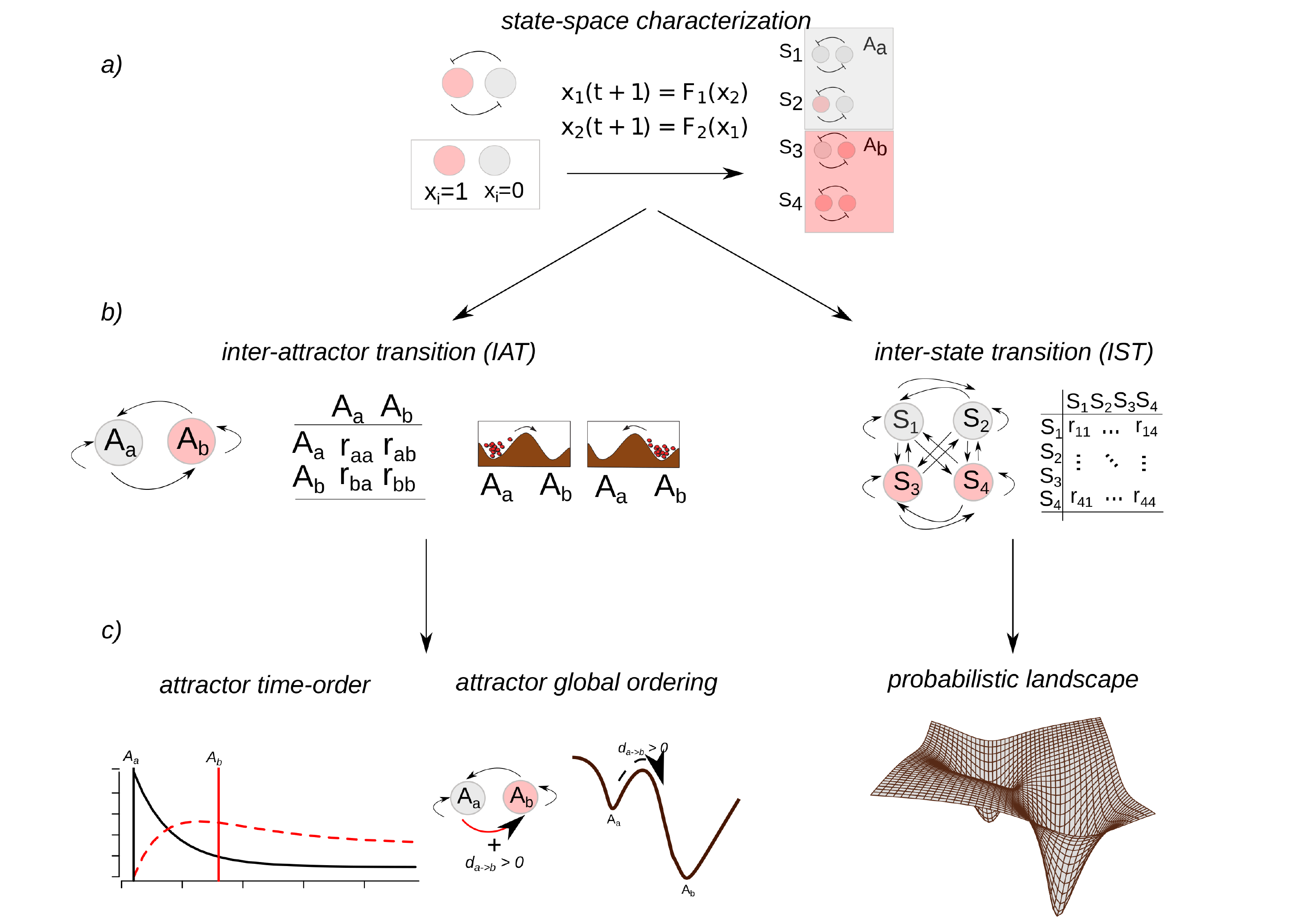}
	\caption{{\small Schematic representations of the general work flow for characterizing the EAL.}}
	\label{fig:hgscores}
\end{figure}

\begin{figure}[p]
	\centering
	\includegraphics[width=150mm]{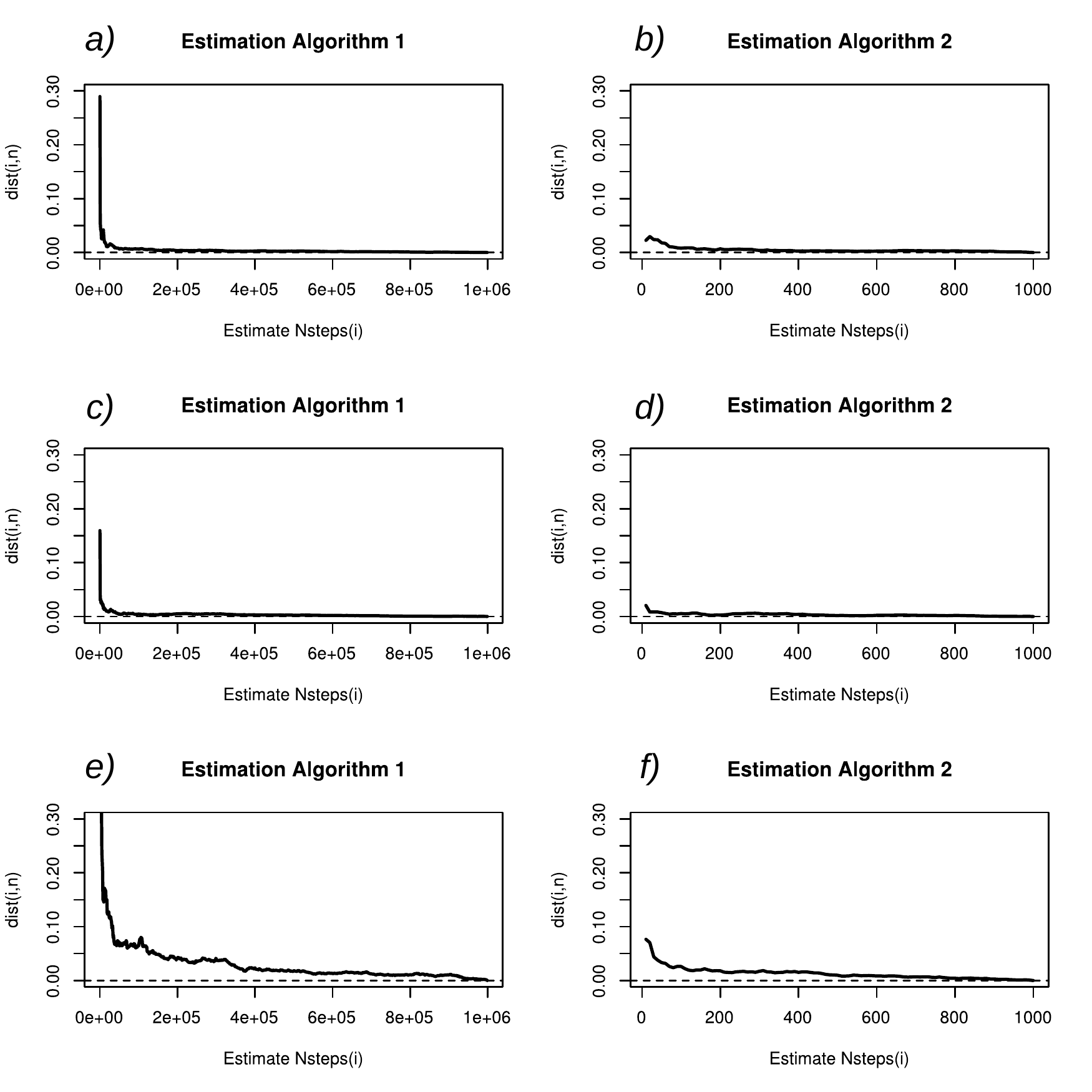}
	\caption{{\small Distance between estimates as a function of parameters $Nsteps$ and $Nreps$.}}
	\label{fig:hgscores}
\end{figure}

\begin{figure}[p]
	\centering
	\includegraphics[width=150mm]{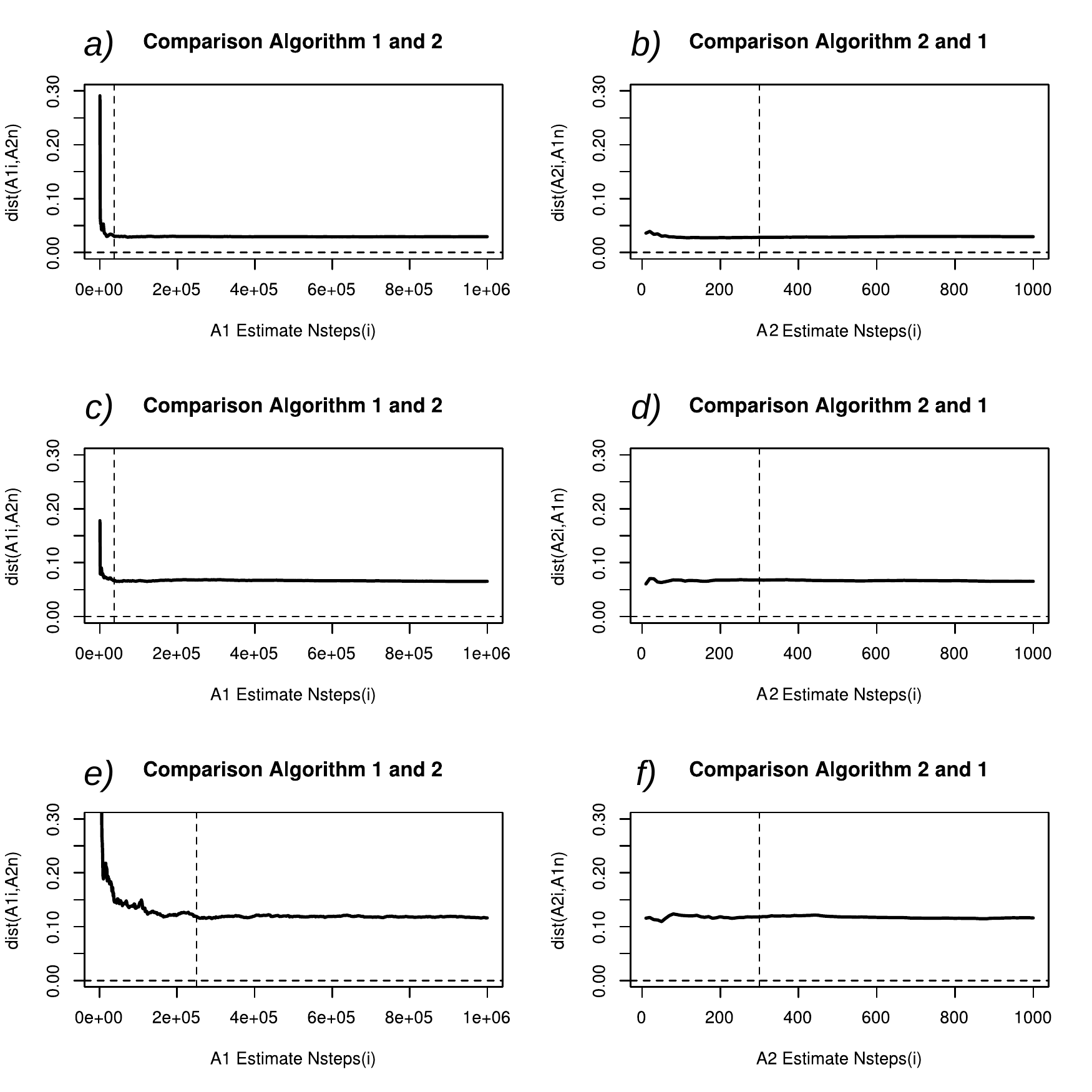}
	\caption{{\small Distance between estimates obtained with Algorithms 1 and 2.}}
	\label{fig:hgscores}
\end{figure}

\begin{figure}[p]
	\centering
	\includegraphics[width=150mm]{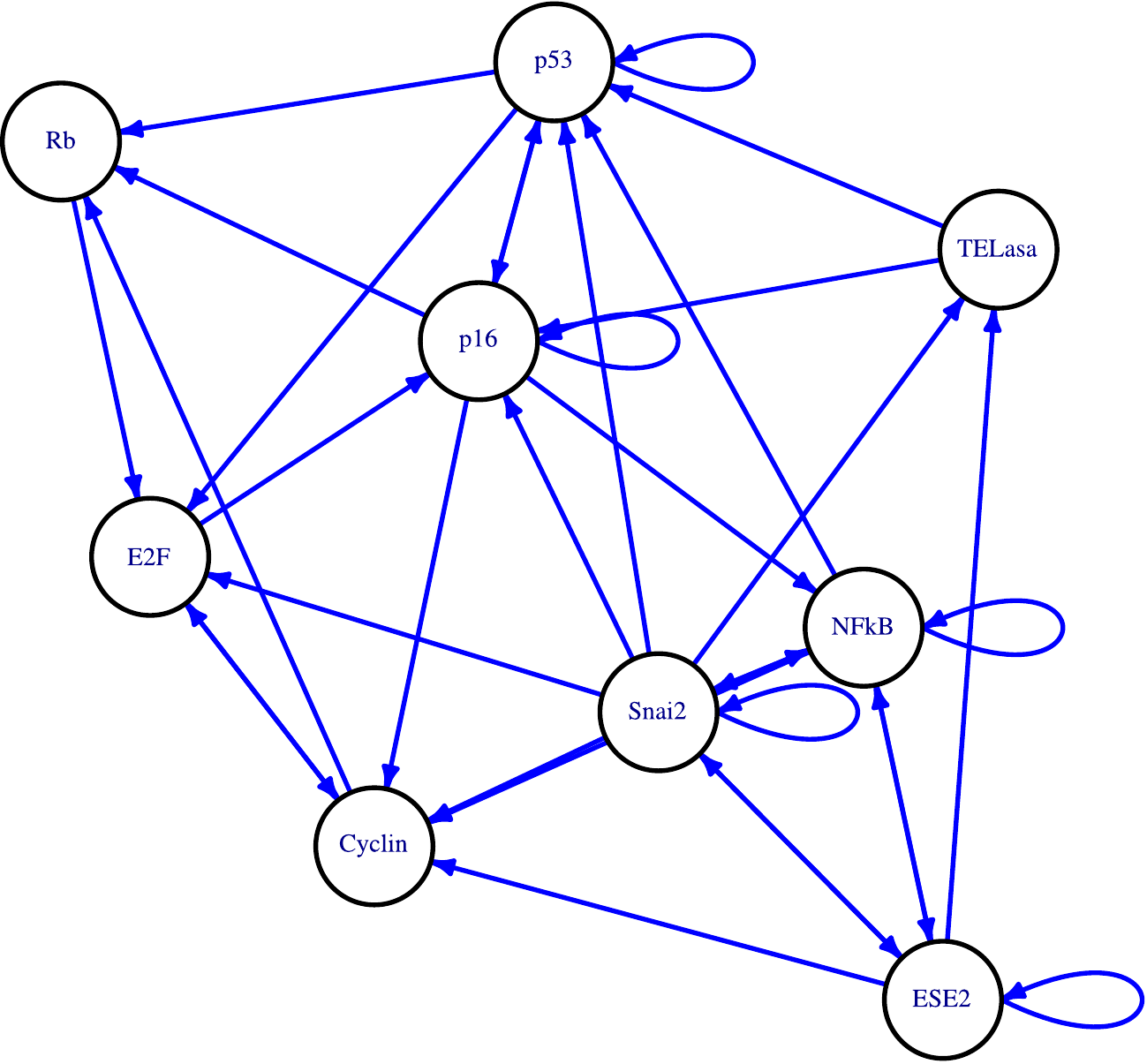}
	\caption{{\small Gene regulatory network for epithelial carcinogenesis.}}
	\label{fig:hgscores}
\end{figure}

\begin{figure}[p]
	\centering
	\includegraphics[width=150mm]{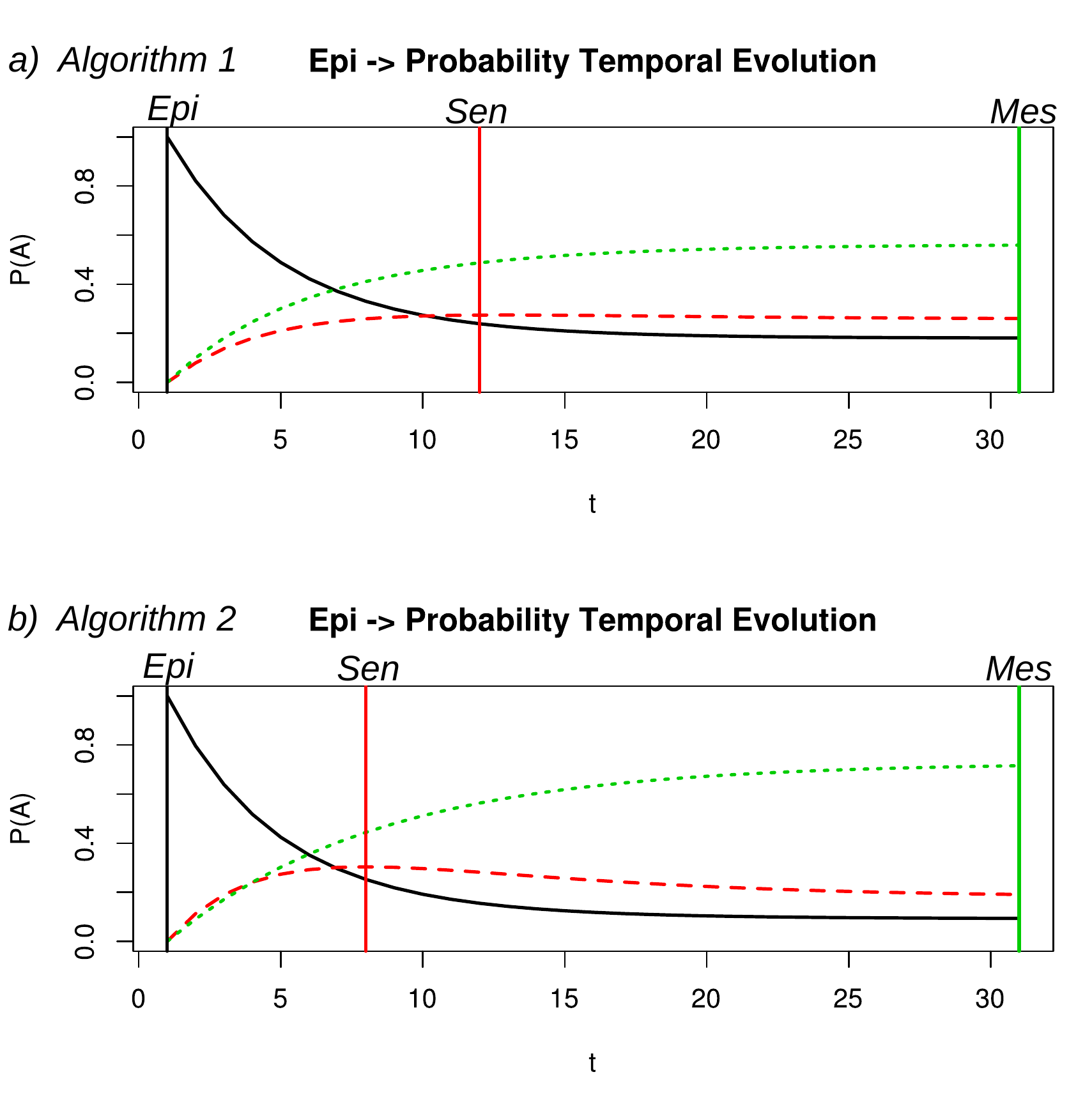}
	\caption{{\small Temporal sequence of cell--fate attainment pattern under the stochastic Boolean GRN model during epithelial carcinogenesis.}}
	\label{fig:hgscores}
\end{figure}

\begin{figure}[p]
	\centering
	\includegraphics[width=150mm]{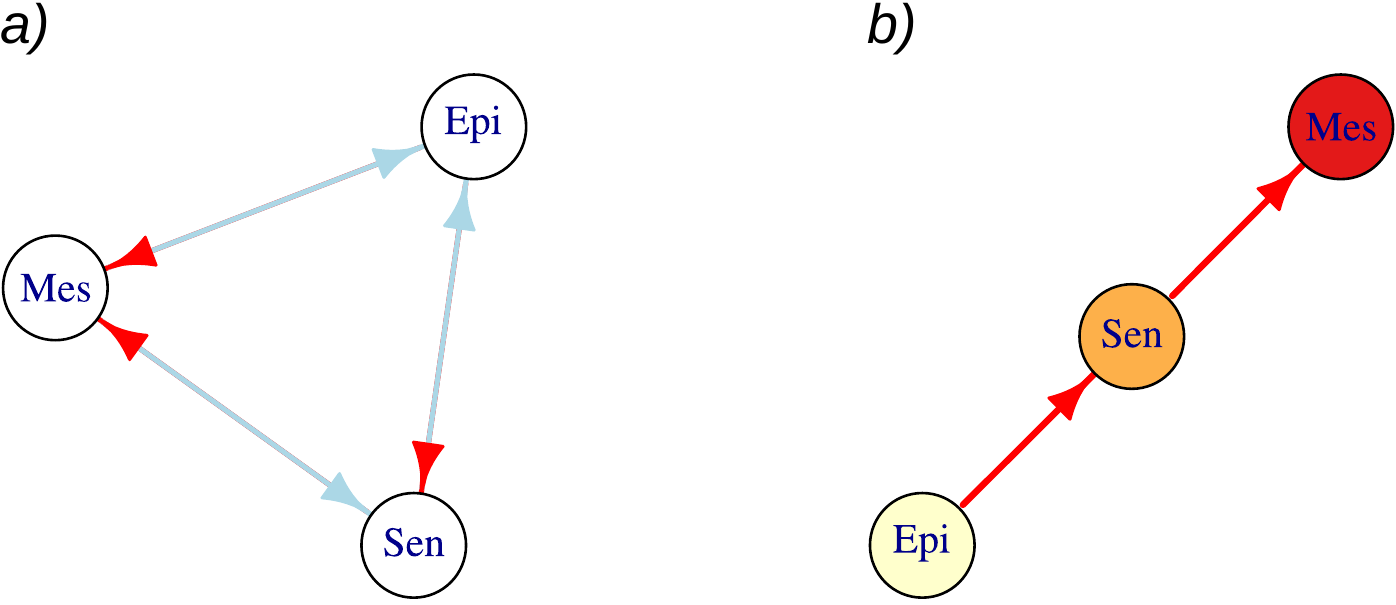}
	\caption{{\small Graph-based representation of attractors transitions.}}
	\label{fig:hgscores}
\end{figure}

\begin{figure}[p]
	\centering
	\includegraphics[width=150mm]{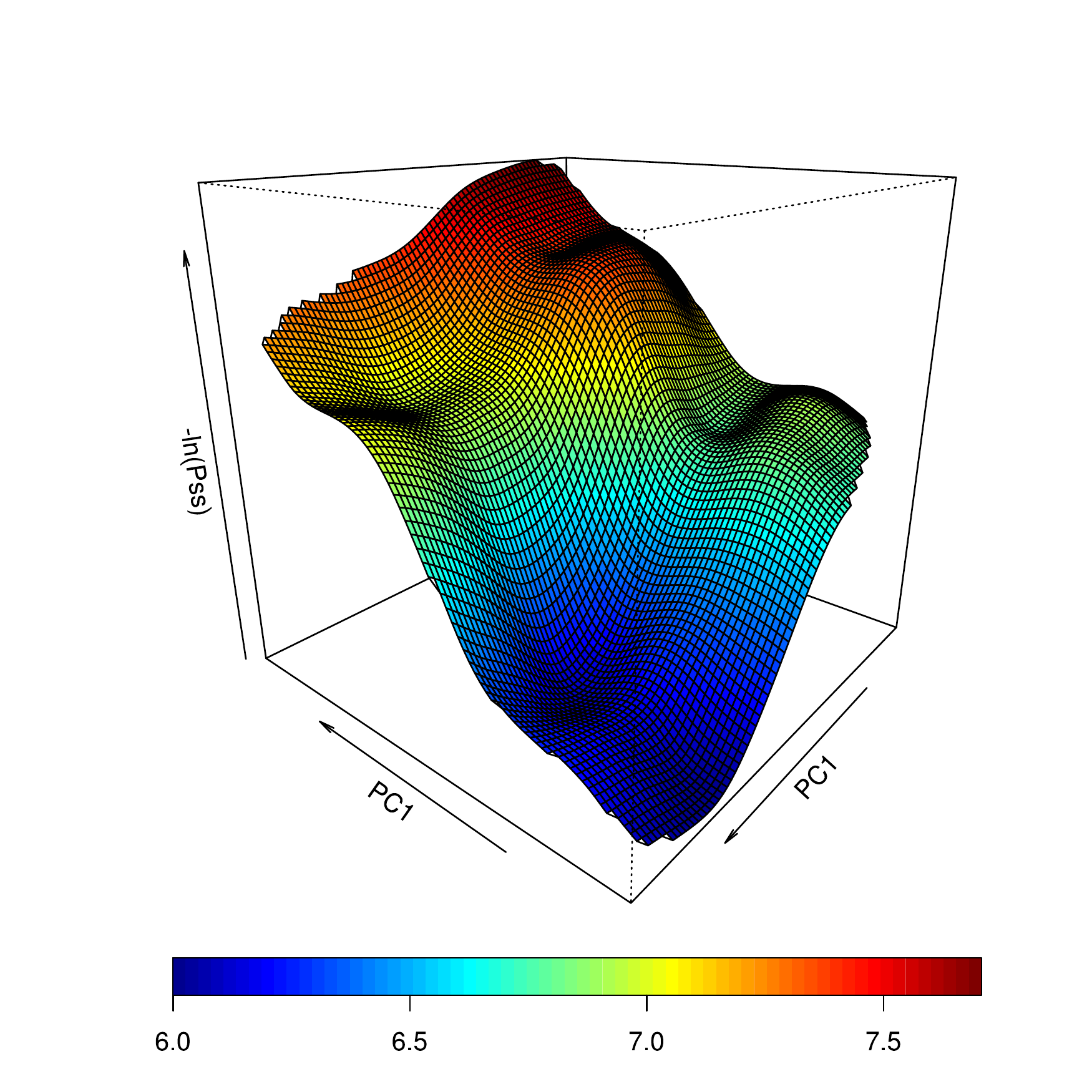}
	\caption{{\small 3D-surface representation of the probabilistic landscape.}}
	\label{fig:hgscores}
\end{figure}

\begin{figure}[p]
	\centering
	\includegraphics[width=150mm]{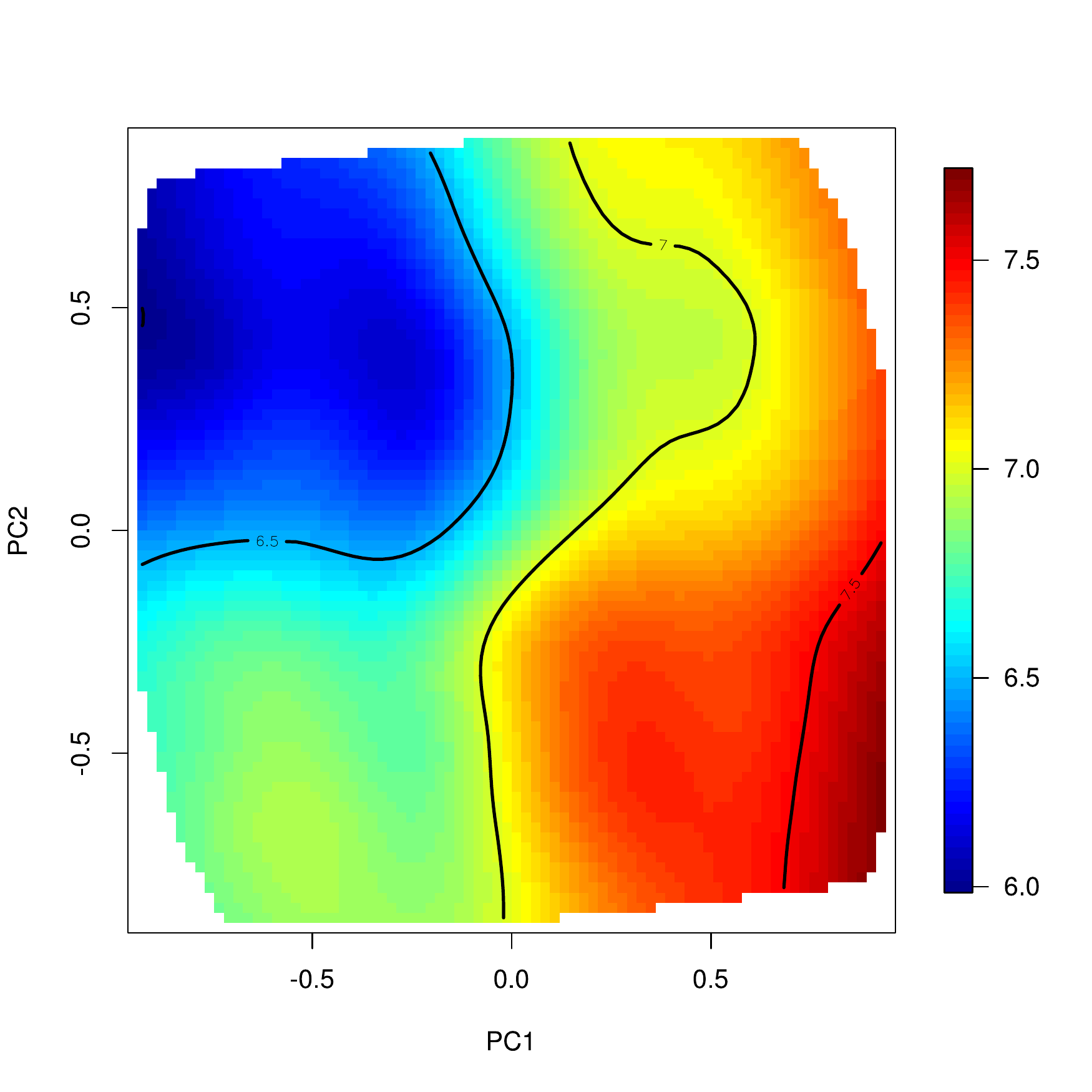}
	\caption{{\small Contour plot representation of the probabilistic landscape.}}
	\label{fig:hgscores}
\end{figure}

\end{document}